\documentclass[12pt]{article}

\def\mco{\multicolumn}
\def\epp{\epsilon^{\prime}}
\def\vep{\varepsilon}
\def\ra{\rightarrow}
\def\ppg{\pi^+\pi^-\gamma}
\def\vp{{\bf p}}
\def\K{{\cal K}}
\def\W{{\cal W}}

\def\ko{K^0}
\def\kb{\bar{K^0}}
\def\al{\alpha}

\def\be{\begin{equation}}
\def\ee{\end{equation}}
\def\bea{\begin{eqnarray}}
\def\eea{\end{eqnarray}}
\def\bt{\begin{tabular}}
\def\et{\end{tabular}}

\newcommand{\go}{\omega}

\newcommand{\f}{\frac}
\newcommand{\p}{\partial}
\newcommand{\half}{\frac{1}{2}}
\newcommand{\ga}{\alpha}

\newcommand{\gd}{\delta}
\newcommand{\gl}{\lambda}

\newcommand{\gvep}{\varepsilon}

\newcommand\un{{{n}}}

\newcommand\um{{{m}}}

\newcommand\ls{\!\!\!\!\!\!\!}

\begin{document}

\begin{center}
{\large\bf Higher Spin Gauge Theories in Any Dimension}
\vglue 0.6  true cm
\vskip0.8cm
{M.A.~Vasiliev}
%\vskip1cm
\vglue 0.3  true cm

%\medskip
Lebedev Physical Institute,
Leninsky prospect 53, 119991, Moscow, Russia
%\medskip
%\vskip2cm
\vskip1.3cm
\end{center}

\begin{abstract}
Some general properties of higher spin gauge theories are
summarized with the emphasize on the nonlinear theories
in any dimension.
\end{abstract}

\newcommand{\bee}{\begin{eqnarray}}
\newcommand{\eee}{\end{eqnarray}}
\newcommand{\nn}{\nonumber}
\newcommand{\lis}{Fort1,FV1,LV}
\newcommand{\hy}{\hat{y}}

%
% Caligraphic letters
%
\def\cala  {{\cal A}} \def\calb  {{\cal B}} \def\calc  {{\cal C}}
\def\cald  {{\cal D}} \def\cale  {{\cal E}} \def\calf  {{\cal F}}
\def\calg  {{\cal G}} \def\calh  {{\cal H}} \def\cali  {{\cal I}}
\def\calj  {{\cal J}} \def\K  {{\cal K}} \def\call  {{\cal L}}
\def\calm  {{\cal M}} \def\caln  {{\cal N}} \def\calo  {{\cal O}}
\def\calp  {{\cal P}} \def\calq  {{\cal Q}} \def\calr  {{\cal R}}
\def\cals  {{\cal S}} \def\calt  {{\cal T}} \def\calu  {{\cal U}}
\def\calv  {{\cal V}} \def\W  {{\cal W}} \def\calx  {{\cal X}}
\def\caly  {{\cal Y}} \def\calz  {{\cal Z}}
%
% Numbers
%
%\def\dl           {\bf}
\def\dl            {\mathbb } \def\complex       {{\dl C}}
\def\naturals      {{\dl N}} \def\rationals     {{\dl Q}}
\def\reals         {{\dl R}} \def\zet           {{\dl Z}} \def\Zet
{${\dl Z}$} \def\zetminuso     {\mbox{$\zet_{\le0}$}} \def\zetplus
{\mbox{$\zet_{>0}$}} \def\zetpluso      {\mbox{$\zet_{\ge0}$}}
%
% LaTeX
%
 \def\comp  {Com\-mun.\wb Math.\wb Phys.}
 \def\nupb  {Nucl.\wb Phys.\ B}
 \def\npbF  {Nucl.\wb Phys.\ B\vypF}
 \def\phlb  {Phys.\wb Lett.\ B}
\def\prd{{\em Phys. Rev.} {\bf D}}

% Standard macros
%

%\usepackage{latexsym}
%\usepackage{colors}
%\begin{document}

\def\mco{\multicolumn} \def\epp{\epsilon^{\prime}}
\def\vep{\varepsilon} \def\ra{\rightarrow}
\def\ppg{\pi^+\pi^-\gamma} \def\vp{{\bf p}} \def\ko{K^0}
\def\kb{\bar{K^0}} \def\al{\alpha}
%\def\W{{\cal W}}
%\def\M{{\cal M}}
%\def\N{{\cal N}}
%\def\K{{\cal K}}
%\def\E{{\cal E}}
%\newcommand{\ggg}{\gamma}

%%%%%%%%%%%%%%%%%%%%%
%%%%%%%%%%%%%%%%%%%%%%%%%%%%%%%%%%%%%%%%%%%%%%%%%%%%%%%%%%%%%%

% \title{{\Large Higher Spin Gauge
%Theories in Any Dimension}} \author{Mikhail Vasiliev}

%\vskip 36pt

%%%%%%%%%%%%%% nomer 1    %%%%%%%%%%%%%%%%%%%%%%

% \thispagestyle{empty}\vskip 2cm \begin{center}

%{\LARGE\it\bf Higher Spin Gauge Theories in Any Dimension}
%Lessons \ from \ open-string \vskip 2pt partition \ functions}

%{\bf\vskip 60pt Mikhail \ Vasiliev \vskip -.2cm {\tiny \it
%Lebedev Institute, Moscow} \small \vskip 40pt {\large \bf Strings
%2004 } \vskip 24pt {\it Paris, June 28, 2004}} \end{center}

%%%%%%%%%%%%%%%%%%%%%%%    nomer2    %%%%%%%%%%%%%%%%%%%%%%%%%%%%%%%%%%%
\section{Introduction}
First  of all, I would like to thank the organizers for the
invitation to talk on higher spin gauge theory
at Strings 2004. Although
a relationship of the higher spin (HS) gauge theory to superstring
theory is not yet completely clear, the
impressive convergency that took place during recent years indicates
that HS theories and Superstring  might be  different faces of
the same fundamental theory to be found.
(For an interesting new argument in the same direction see
\cite{MB} and the talk of M.Bianchi at this conference).
\subsection{Symmetric Massless Free Fields}
Simplest free HS gauge fields
are so called totally symmetric fields. They can be
described by rank $s$ totally symmetric tensors
$\varphi_{n _1\ldots n_s}(x)$ subject to the
double-tracelessness condition $\varphi^m{}_{m}{}^k{}_
 {k\,n_5\ldots n_s}(x)=0$ \cite{CF} ($m,n,\ldots =0,\ldots , d-1$).
The Abelian HS gauge symmetries $\delta\varphi_{n_1\ldots n_s}(x)
=\partial_{\{n_1}\varepsilon_{n_2\ldots n_{s}\}}(x) $ with rank
$s-1$ symmetric traceless gauge parameters $\varepsilon_{n_1\ldots
n_{s-1}}(x)$ ($\varepsilon^{r}{}_{r n_3\ldots n_{s-1}}(x)=0$)
leave invariant the quadratic action \bee \ls\ls&{}&S^s=\frac{1
}{2 }(-1)^s \int d^d x\,\{\partial_n\varphi_{m_1\ldots
m_s}\partial ^n\varphi ^{m_1 \ldots m_s}   \\
\ls\ls\ls\ls&-&\frac{s(s-1)}{2}\partial_n\varphi^r {}_{r m_1 \ldots
m_{s-2}}\partial^n\varphi^k{}_k{}^{m_1\ldots m_{s-2}}
+s(s-1)\partial_n\varphi^r{}_{r m_1\ldots
m_{s-2}}\partial_k\varphi ^{n k m_1\ldots m_{s-2}}\nn\\
\ls\ls\ls\ls&-&s\partial_n\varphi^n{}_{m_1\ldots m_{s-1}}\partial _r\varphi^{r
m_1\ldots m_{s-1}} -
\frac{s(s-1)(s-2)}{4}\partial_n\varphi^r{}_r{}^n{}_{m_1\ldots m
_{s-3}}\partial_k\varphi^t{}_t{}^{k m_1\ldots m_{s-3}}\}\,.\nn
\eee
This action describes a spin $s$ massless field \cite{CF} and
generalizes the spin 1 Maxwell action and
spin 2 (linearized) Einstein action to any integer spin.
The key question is what is a
fundamental unifying symmetry principle underlying HS gauge fields.
Even at the free field level, the existence of elegant
metric-like \cite{metr} and frame-like  \cite{frame, framed}
``geometric formulations" indicates that there must
be some deep reason for HS theories to exist.

\subsection{Higher Spin Problem}%{\vskip -.5cm}

%%%%%%%%%%%%%% nomer 3    %%%%%%%%%%%%%%%%%%%%%%
%%%%%%%%%%%%%%%%%%%%%%%%%%%%%%%%%%%%%%%%%
The problem is to find  a nonlinear HS gauge  theory such that
it has
%%%%%%%%%%%%%%%%%%%%%%%%%%%%%%%%%%%%%%%%%%%%%%%%%%
\begin{itemize}
\item
Correct free field limit
\item
Unbroken  HS  gauge
symmetries
\item
Non-Abelian global HS
symmetry  of  a  vacuum  solution
\end{itemize}
The first condition demands the theory to be free of ghosts,
that is  to  be equivalent to the Fronsdal theory for the case of
totally symmetric fields. The third condition avoids trivial
possibility of Abelian interactions built of Abelian gauge
invariant HS field strengths like nonlinear terms built
of higher powers of the spin 1  Abelian field strength
instead of Yang-Mills interactions.

The HS problem is of interest on its own right. An additional
stringy motivation is, in the first place, that
it is tempting to interpret
massive  HS modes in Superstring as resulting from  breaking
 of  HS  gauge  symmetries.
In that case, superstring should
exhibit higher symmetries in the high-energy limit as was
argued long ago by Gross  \cite{Gross}.
A more recent argument came from the $AdS/CFT$ side
after it was realized  \cite{Su} that
HS symmetries should be unbroken in the
Sundborg--Witten limit
$\gl=g^2N\rightarrow 0\,,$
$l^2_{str}\Lambda_{AdS}\rightarrow \infty$
just because the boundary conformal theory becomes free.
A dual string theory in the highly curved $AdS$ space-time
is therefore going to be a HS theory.

\subsection{Difficulties}

Although the formulation of the HS problem may look
rather flexible, the conditions
on the HS interactions is so hard to satisfy
that many  believed it
admits no solution at all. One difficulty
is due to the $S$-matrix argument
{\it a la} Coleman-Mandula \cite{CM} that, if $S$-matrix
has too many symmetries carrying nontrivial representations
of the Lorentz symmetry as HS symmetries do, then $S=I$, i.e.
there is no interactions.

Another one is the HS-gravity interaction problem as was originally
pointed out by
 Aragone and  Deser in \cite{AD}.
The point is that covariantization of derivatives
$
\partial_n \rightarrow D_n=\partial_n -\Gamma_n      \,,
$ $ \gd \varphi_{nm\dots}\rightarrow D_n \varepsilon_{m\dots} $
changes the situation drastically because they do not commute, $
[D_n\,, D_m]={\cal R}_{nm}\dots $,
 if the Riemann tensor ${\cal
R}_{nm,pq}$ is nonzero. As a result, the variation of the
covariantized HS action under covariantized HS gauge theories is
not any longer zero \be \label{prob} \delta S_s^{cov} =\int {\cal
R}_{\ldots}(\varepsilon_{\ldots} D\varphi_{\ldots})\neq 0\,\qquad
?! \ee Most important is that the Weyl tensor part of the Riemann
tensor contributes to this variation for $s>2$,
 which contribution
seems to be hard to compensate by any modification of the
action and/or field transformations\footnote{Recall that,
for spin 3/2, analogous terms can be compensated by the
modification of the transformation of the
metric tensor under local SUSY transformation
because only Ricci tensor contributes in this case,
that opens a way towards supergravity.}.

\subsection{Resolution}
Despite the difficulties with HS interactions,
in the important works \cite{BBB,BBD}
it was shown that some consistent (i.e., gauge invariant)
interactions of HS gauge
fields with matter fields and with themselves do exist
at least at the cubic level
\be
\label{act}
\ls S=S^2+S^3+\dots \qquad S^3=\sum_{p,q,r}(D^p
\varphi)(D^q \varphi)(D^r \varphi){\ell}^{\,p+q+r+\half
d-3}\,,
\ee
where $\ell$ is a parameter of dimension of length.
These authors discovered that interactions
of HS fields contain higher space-time derivatives:
the higher interacting spins are, the more (but finite)
number of derivatives appear in their interactions.

Another important observation was \cite{FV} that
the situation with HS interactions and, in particular, with
HS-gravitational interactions, improves once the
problem is reconsidered in the $(A)dS$ background.
The dimensionful parameter $\ell$ then identifies with the radius of
$(A)dS$ space
$
 {\ell}=\Lambda^{-\half}=R_{AdS}\,.
$
The key difference  between flat and $(A)dS$ background is that,
in the latter case, background covariant derivatives do not commute
$
[D_n\,, D_m ] \sim \Lambda \sim O(1)\,.
$
This has an important consequence that the terms of
different orders in derivatives in the action (\ref{act})
do talk to each other. As a result, there exists \cite{FV} such a
unique (modulo field redefinitions)
combination of higher derivative interaction terms that
a contribution from their gauge variation
cancels the problematic terms (\ref{prob}) of the
flat space analysis. Since the coupling constants of the
interaction terms contain positive powers of the
$(A)dS$ radius, they blow up in the flat limit in
agreement with the flat space no-go results. In $(A)dS$ space,
the no-go arguments do not work (in particular, no $S$-matrix).
HS theories suggest deep
analogy between the $AdS$ scale and string length scale
$
{\ell} \sim \sqrt{\alpha^\prime}\,,
$
although the precise identification is far from being clear
at the moment. Note that the role of $(A)dS$ background
in HS theories was realized years before the discovery of
the $AdS/CFT$ correspondence \cite{ADSCFT}.

\section{HS Fields as Gauge Connections}
\label{HS fields as gauge connections}
It is well-known  that gauge fields of supergravity result from
 gauging the SUSY algebra:
%\vskip 15pt
\begin{center}\begin{tabular}{l l l l} {\bf
$o(d-1,2)$}&&{\bf $o(N)$}&\\
%\hspace{6cm}&\hspace{6cm}&\hspace{6cm}&\hspace{6cm}\\
 \rule{0pt}{10pt}$T^{A\,B}$&$Q^p_\ga$\rule{30pt}{0pt}
 &$t^{p\,q}$\rule{30pt}{0pt}&
{\bf\small $A,B,\dots=0,\dots d$},\qquad { \bf\small $p,q,\dots=1,\dots  N$}\\
%&&&\\
 \rule{0pt}{10pt}$\go_n{}^{A\,B}$&$\Psi_{n}{}^p_\ga$&
 $A_n{}^{p\,q}$& { \small $n=0,\dots  d-1\qquad\quad\quad \ga\quad
\mbox{is spinor index}$}
\end{tabular}
\end{center}
In particular, $s=2$ gravitational field is described in
supergravity by the frame field $e_\un^a$ which along with the
Lorentz connection $\go_\un^{ab}$ can be interpreted \cite{MMSW}
as components of the gauge field $\go_\un^{AB}$
of the $AdS_d$ algebra $o(d-1,2)$\\
$\rule{90pt}{0pt}
 g_{\un\um}\longrightarrow e_\un^a\longrightarrow
\{e_\un^a,\go_\un^{ab}\}\rule{0pt}{25pt}
 \longrightarrow \go_\un^{AB} \qquad
$
{ \begin{picture}(10,23)
%stolbik 2
{\linethickness{.30mm}
\put(10,20){\line(1,0){10}}%
\put(10,10){\line(1,0){10}}%
\put(10,0){\line(1,0){10}}%
\put(10,0){\line(0,1){20}}%
\put(20,0.0){\line(0,1){20 }}} \end{picture}}
%\smallskip\rule{0pt}{15cm}
% \rule{435pt}{0pt}{\bf\small MacDowell, Mansouri\, (1978)}\rule{0pt}{35pt}\\
%\rule{520pt}{0pt}{\bf\small Stelle, West \,\,(1980)}\\

Analogously, a totally symmetric field $\varphi_{\un_1\ldots \un_s}$
in the Fronsdal formulation
admits an equivalent description in terms of the gauge 1-form
$\omega_\un^{A_1\ldots A_{s-1},B_1\ldots B_{s-1}}$\cite{framed}
\be
 \varphi_{\un_1\ldots \un_s} \rightarrow e_\un{}^{a_1\ldots
a_{s-1}}
 \rightarrow \{e_\un{}^{a_1\ldots a_{s-1}},\go_\un{}^{a_1\ldots a_{s-1},b_1
 \ldots
 b_{s-1}}\}
 \rightarrow
\omega_\un^{A_1\ldots A_{s-1},B_1\ldots
B_{s-1}}\,,
\ee
which takes values in the irreducible representation of the $AdS_d$
algebra $o(d-1,2)$ depicted by the (traceless) rectangular two-row
Young tableau\\
  \begin{tabular}{ll}
$%\left.
\begin{array}{l}
\qquad\qquad \omega_\un^{\{A_1\ldots A_{s-1},A_s\}\,B_2\ldots B_{s-1}}=0\\
\qquad\qquad \omega_\un^{A_1\ldots A_{s-3}\,C\,}{}_{C,}{}^{B_1\ldots
B_{s-1}}\,=\,0 \end{array} %\right\}
$&\rule{0pt}{30pt}\qquad\qquad\qquad{
\begin{picture}(80,25)(0,5)
%2palki s-1
{\linethickness{.250mm}
\put(00,00){\line(1,0){80}}%
\put(00,10){\line(1,0){80}}%
\put(00,20){\line(1,0){80}}%
\put(00,00){\line(0,1){20}}%
\put(10,00.0){\line(0,1){20}} \put(20,00.0){\line(0,1){20}}
\put(30,00.0){\line(0,1){20}} \put(40,00.0){\line(0,1){20}}
\put(50,00.0){\line(0,1){20}} \put(60,00.0){\line(0,1){20}}
\put(70,00.0){\line(0,1){20}} \put(80,00.0){\line(0,1){20}} }
\put(20,24) {{ \small \bf ${s-1}$}} \put(10,-16){\small\bf
$o({d-1},2)$} \end{picture}}\\ \end{tabular}
\\
\\

Let an $AdS$ vector $V_A$ define the Lorentz subalgebra
$o(d-1,1)\subset o(d-1,2)$ as its stability subalgebra.
The simplest choice is $V_B=\gd_B^{\hat d}$ where
$ {\hat d} $ denotes the $(d+1)^{th}$
 Lorentz invariant direction of an $AdS$ vector.
The HS dynamical frame-like field is then identified with the
components of the HS connection with a maximal possible number
of $AdS_d$ vector components along the extra direction $V_A$
\be
e_\un{}^{a_1\ldots a_{s-1}}=\omega_\un^{a_1\ldots
a_{s-1}}{}_,{}^{B_1\ldots B_{s-1}} V_{B_1}\ldots
V_{B_{s-1}}\,.
\ee
Analogously to the spin 2 metric case,  Fronsdal field is
the totally symmetric part of the frame field
$
 \varphi_{\un_1\un_2\ldots \un_s}=
e_{\{\un_1\,,\,{\un_2\ldots \un_s}\}}\,.
$
Generalized Lorentz connections identify with those components
of the connection that carry more Lorentz indices
\be
\go_\un{}^{a_1\ldots
a_{s-1}}{}_,{}^{b_1\ldots b_{t}}=
\omega_\un^{a_1\ldots a_{s-1}}{}_,{}^{b_1\ldots
b_{t}\,B_{t+1}\ldots B_{s-1}} \,V_{B_{t+1}}\ldots
V_{B_{s-1}}\,.
\ee
Upon resolving appropriate torsion-like constraints \cite{framed},
the generalized Lorentz connections are expressed through
 derivatives of the frame-like field\hfil\\
$
\go_\un{}^{a_1\ldots
a_{s-1}}{}_,{}^{b_1\ldots b_{t}}\sim
 \left( \f{1}{\sqrt{\Lambda}}\f{\p}{\p
x}\right)^{t}(e)
$
so that every additional Lorentz index brings one derivative
along with one power of $\ell = \f{1}{\sqrt{\Lambda}}$.
In this formalism, the higher derivatives of HS interactions,
as well as negative powers of the cosmological constant,
result from the dependence of the nonlinear terms in the
HS actions on the higher generalized Lorentz connections.

\section{Higher Spin Algebra $hu (1|2\!\!:\!\![d\!-\!1,2])$}

The structure of HS gauge connections suggests that they result
from gauging a HS algebra that contains the $AdS_d$ algebra $o(d-1,2)$
as a subalgebra and decomposes under adjoint action of the latter
into a sum of representations described by traceless two-row Young
tableaux. In other words, generators $T_{A_1\ldots
A_{n},B_1\ldots B_{n}}$ of a HS algebra should satisfy the conditions
$
T_{\{A_1\ldots
A_{n},A_{n+1}\}B_2\ldots B_{n}}=0$ and
$T^C{}_{C\,A_3\ldots A_{n},B_1\ldots B_{n}}=0\,.$
Such an algebra was originally found by
Eastwood \cite{East} as conformal HS algebra of a scalar field
in $d-1$ dimensions. For our purpose it is most convenient to
use its oscillator realization suggested in
\cite{d}.

Namely we introduce a canonical pair of $AdS$
vectors $Y^A_i$,
\be
\label{osc}
[Y_i^A\,,\,Y_j^B]_*\,=\,\epsilon_{ij}\eta^{AB}\,,
\ee
where $\eta_{A\,B}=\eta_{B\,A}$ is the $AdS_d$ invariant metric
and $\epsilon_{i\,j}=-\epsilon_{j\,i}\quad i,j =1,2$ is the
$sp(2)$ invariant form. Here we use the star product
notation for the oscillator algebra defined by the relations
(\ref{osc}) (for its precise definition see eq.(\ref{star})
for $Z$-independent functions).
The bilinear combinations of oscillators, $T^{AB}$ and
$t_{ij}$,
 \be
\label{sp2}
T^{A,B}=-T^{B,A}\,=\,\half \{Y_i^A\,,\,Y_j^B\}_*
 \epsilon^{ji}\,,\qquad
 t_{i\,j}=t_{j\,i}\,=\,\half
\{Y_i^A\,,\,Y_{j\,A}\}_* \,,
\ee
form, respectively, the  $o(d\!-\!1,2 )$ generators, which rotate
$AdS_d$ vector indices $A,B$, and $sp(2)$ generators,
which rotate symplectic indices $i,j$.
They commute to each other,
$
[t_{i\,j}\,,\,T^{A\,B}]_*\,=\,0\,,
$
thus being Howe dual.
Let us note that the $sp(2)$ plays here a role analogous
to that in the description of dynamical models
in the conformal framework (two-time physics)
\cite{mar,bars}.

Now it is easy to define the simplest HS algebra
$hu(1|2\!\!:\!\![d\!-\!1,2])$ by virtue of a sort of
Hamiltonian reduction. First one considers the Lie
algebra of functions of oscillators with the Lie bracket
 $[f(Y)\,,\,g(Y)]_*$. Then one
considers its subalgebra $S$
spanned by $sp(2)$ invariants
\be
\label{centr}
[f(Y)\,,\,t_{ij}]_*\,=\,0
\ee
and next its quotient $S/I$ where
the ideal $I$ is spanned by the elements
proportional to the $sp(2)$ generators i.e.,
$\{f\in I :\quad
f(Y)\,=\,t^{i\,j}*f_{l\, i\,j}\,\,=\,f_{r\, i\,j}*t^{i\,j}\,\sim\,0\,.\}$
The algebra $S/I$ we call
 $hu(1|2\!\!:\!\![d\!-\!1,2])$ (upon imposing appropriate
 reality conditions \cite{d}).

%\section{Gauging HS Algebra}

The gauge fields of $hu(1|2\!\!:\!\![d\!-\!1,2])$ are
\be
\label{gauge}
 \go(Y|x)=\sum_{n=0}^\infty dx^\um \go_{\um\,A_1\ldots
A_{n},B_1\ldots  B_{n}}(x) Y_1^{A_1}\ldots Y_1^{A_{n}} \,
Y_2^{B_1}\ldots Y_2^{B_{n}}\,.
\ee
It is easy to see that the condition (\ref{centr})
(which can be written in the covariant form
$
Dt_{i\,j}\,=\,d t_{i\,j}  + [\go\,,\,t_{i\,j}]_*=0
$
taking into account $d t_{i\,j}=0$)
imposes the Young properties\rule{0pt}{10pt}
\begin{center}
 \bt{lr} $\go_{\um\,\{A_1\ldots
A_{n},A_{n+1}\} B_2\ldots B_{n}}=0
 \rule{0pt}{15pt}$
& {\begin{picture}(80,20)(0,5)
%2palki s-1
{\linethickness{.250mm}
\put(00,00){\line(1,0){80}}%
\put(00,10){\line(1,0){80}}%
\put(00,20){\line(1,0){80}}%
\put(00,00){\line(0,1){20}}%
\put(10,00.0){\line(0,1){20}} \put(20,00.0){\line(0,1){20}}
\put(30,00.0){\line(0,1){20}} \put(40,00.0){\line(0,1){20}}
\put(50,00.0){\line(0,1){20}} \put(60,00.0){\line(0,1){20}}
\put(70,00.0){\line(0,1){20}} \put(80,00.0){\line(0,1){20}} }
\put(32,21.2){  ${n}$}%\put(32,-11.2){  $o({d-1},2)$}
\end{picture}} \et
\end{center}
(including the property that the r.h.s. of (\ref{gauge})
contains equal numbers of oscillators $Y^A_1$ and $Y^A_2$).
According to (\ref{sp2}), the factorization over the
terms proportional to $t_{ij}$
is equivalent to factorization of traces of the gauge
field components in (\ref{gauge}) that gives rise
precisely to the set of gauge fields associated with
different spins as explained in section
\ref{HS fields as gauge connections}.

The non-Abelian HS field strength is
\be
\label{HSCUR}
R = d \go(Y|x) +\go(Y|x)\wedge * \,\go(Y|x)\,,
\ee
where terms on the r.h.s., which take values in the ideal
$I$, are factored out. The infinite-dimensional HS algebra
contains the maximal finite-dimensional subalgebra
$o(d-1,2) \oplus u(1)$ spanned by bilinears in oscillators
and constants, respectively.
Different spins correspond to homogeneous polynomials
$\go(\mu Y|x)=\mu^{2(s-1)}\go(Y|x) $. The gauge fields of
$o(d-1,2) \oplus u(1)$
carry spin 2 and spin 1 respectively. That
$o(d-1,2) \oplus u(1)$ is a maximal finite-dimensional subalgebra
of $hu(1|2\!\!:\!\![d\!-\!1,2])$ is a consequence
of the fact that the commutator of degree $p$ and degree $q$
polynomials of oscillators gives a degree $p+q-2$ polynomial.
For example, if spin 3 associated with degree 4 polynomials
in oscillators appears, polynomials of all higher degrees appear
in the closure of its generators. Thus, beyond the barrier of spin 2,
the systems of HS fields are necessarily
infinite. Let us note that there exists a
generalization of the nonlinear HS gauge theory to the case
with HS gauge connection carrying matrix indices $\go\rightarrow
\go_q{}^p(Y|x)$ \quad $p,q=1,\dots,n$ so that
the spin 1 Yang-Mills algebra is promoted to
$u(n)$ (HS models with the Yang-Mills gauge algebras
$o(n)$ and $usp(n)$ also exist \cite{d}).

\section{Lower Spin Examples}

To illustrate the idea of the approach that allows us to
formulate nonlinear HS dynamics let us start with lower spin
examples.

The nonlinear $s=2$ equations are equivalent to zero-torsion condition
$R^a=0$ together with the Einstein equation in the form
\be
\label{ein}
R^{ab}=e_c{} \wedge e_d \,\,C^{ac},{}^{bd}\,,
\ee
where $e$ is the frame 1-form and $C^{ac},{}^{bd}$ has algebraic
properties of the Weyl tensor, i.e. it is traceless and
symmetrization over three indices gives zero\footnote{For the future
convenience we use symmetric basis with
$C^{ac},{}^{bd}=C^{ca},{}^{bd}=C^{ac},{}^{db}$. The
relationship with the standard antisymmetric Weyl tensor
$\widetilde{C}^{[ac]},{}^{[bd]}$ is $C^{ab},{}^{cd}=
\half\left (\widetilde{C}^{[ac]},{}^{[bd]}
+ \widetilde{C}^{[bc]},{}^{[ad]} \right )$.}. The equation
(\ref{ein}) tells us that $C^{ac},{}^{db}$ is indeed the Weyl
tensor and, because it is traceless, that the Ricci tensor is
zero. Bianchi identities then imply at the linearized level
that non-zero components of order $k$ derivatives of
the Weyl tensor
$\p_{n_1}\ldots \p_{n_k} C_{a_1 a_{2}, b_1 b_2}$ form
 Lorentz tensors
$C_{c_1\ldots c_{k+2}, d_1 d_2}$ described by the Young tableaux
\quad{\begin{picture}(80,35)(0,5)
%2palki s-1
{\linethickness{.250mm}
\put(00,00){\line(1,0){20}}%
\put(00,10){\line(1,0){80}}%
\put(00,20){\line(1,0){80}}%
\put(00,00){\line(0,1){20}}%
\put(10,00.0){\line(0,1){20}} \put(20,00.0){\line(0,1){20}}
\put(30,10.0){\line(0,1){10}} \put(40,10.0){\line(0,1){10}}
\put(50,10.0){\line(0,1){10}} \put(60,10.0){\line(0,1){10}}
\put(70,10.0){\line(0,1){10}} \put(80,10.0){\line(0,1){10}} }
\put(12,23.2){\small \bf ${k+2}$}%\put(32,-11.2){  $o({d-1},2)$}
\end{picture}}
. Einstein equations imply that all $C_{a_1\ldots a_{k+2}, b_1 b_2}$ are
traceless.

In terms of the quantities $C_{c_1\ldots c_{k+2}, d_1 d_2}$,
consequences of
the linearized $s=2$ equations in flat space can be written
in the form of covariant constancy conditions
\be
dC_{a_1\ldots a_{l}, b_1 b_2}=e_0^c\left( l C_{a_1\ldots a_lc,
b_1 b_2} +2 C_{a_1\ldots a_l\{b_1, b_2\} c}\right)\,,
\ee
where $e^a_0$ is the flat Minkowski frame $e^a_0 =dx^a$.

Analogously, $s=1$ Maxwell equations can be reformulated as
\be
F=e_0^c{} \wedge e^d_0 \,\,C_{c},{}_{d}\,,
\ee
\be
dC_{a_1\ldots a_{l}, b }=e_0^c\left( (l+1) C_{a_1\ldots a_{l}c, b} +
C_{a_1\ldots a_{l}b, c}\right)
\ee
with the 0-forms $C_{a_1\ldots a_{l}, b }$ described by the
traceless Young tableaux
\,\,\,\,{\begin{picture}(70,23)(0,2.50)
%2palki k+1,1
{\linethickness{.250mm}
\put(00,00){\line(1,0){10}}%
\put(00,10){\line(1,0){60}}%
\put(00,20){\line(1,0){60}}%
\put(00,00){\line(0,1){20}}%
\put(10,00.0){\line(0,1){20}} \put(20,10.0){\line(0,1){10}}
\put(30,10.0){\line(0,1){10}} \put(40,10.0){\line(0,1){10}}
\put(50,10.0){\line(0,1){10}} \put(60,10.0){\line(0,1){10}}
%\put(70,10.0){\line(0,1){10}}% \put(80,10.0){\line(0,1){10}} }
\put(22,23.2){\small\bf ${k+1}$}}%\put(32,-11.2){  $o({d-1},2)$}
\end{picture}}\\ %\qquad
%\vskip-30pt

$s=0$ Klein-Gordon dynamics is reformulated in the form
 \be
\label{s0}
dC_{a_1\ldots a_{k} }=e_0^c (k+2) C_{a_1\ldots a_{k}c}
\ee
in terms of symmetric traceless 0-forms $C_{a_1\ldots a_{k} }$,
which parametrize all on-mass-shell nontrivial
combination of derivatives of the scalar field $C$ and are
described by the Young tableaux
\,\,\,\,{\begin{picture}(70,18)(0,12)
%1palki s-1
{\linethickness{.250mm}
%\put(00,00){\line(1,0){20}}%
\put(00,10){\line(1,0){70}}
\put(00,20){\line(1,0){70}}%
\put(00,10){\line(0,1){10}}%
\put(10,10){\line(0,1){10}}\put(20,10){\line(0,1){10}}
\put(30,10.0){\line(0,1){10}} \put(40,10.0){\line(0,1){10}}
\put(50,10.0){\line(0,1){10}} \put(60,10.0){\line(0,1){10}}
\put(70,10.0){\line(0,1){10}}% \put(80,10.0){\line(0,1){10}} }
\put(22,21.2){\small\bf ${k}$}}%\put(32,-11.2){  $o({d-1},2)$}
\end{picture}}\,.

This formulation extends naturally to any spin $s$
described by ``Weyl 0-forms"
$C_{a_1\dots a_{s+k}\,,\,b_1\dots b_s}$ with the symmetry
properties of the traceless Young\\\\
tableaux
\quad{\begin{picture}(80,25)(0,0)
%2palki s+k,s
{\linethickness{.350mm}
\put(00,00){\line(1,0){50}}%
\put(00,10){\line(1,0){80}}%
\put(00,20){\line(1,0){80}}%
\put(00,00){\line(0,1){20}}%
\put(10,00.0){\line(0,1){20}} \put(20,00.0){\line(0,1){20}}
\put(30,0.0){\line(0,1){20}} \put(40,0.0){\line(0,1){20}}
\put(50,0.0){\line(0,1){20}} \put(60,10.0){\line(0,1){10}}
\put(70,10.0){\line(0,1){10}} \put(80,10.0){\line(0,1){10}} }
\put(12,22.2){\small\bf  ${s+k}$}\put(12,-10.2){\small\bf  $s$}
\end{picture}}\,.
The meaning of the set of 0-forms
$C_{a_1\dots a_{s+k}\,,\,b_1\dots b_s}$ \\
\\ is that they form a basis
in the space of gauge invariant on-mass-shell nontrivial
derivatives of a massless field under consideration. As a result,
the space
of $C_{a_1\dots a_{s+k}\,,\,b_1\dots b_s}(x)$ at any given
$x$ is analogous (in fact, dual by  a nonunitary Bogolyubov transform)
to the space ${\cal H}$ of spin $s$ single-particle states.
Thus, Weyl 0-forms are sections of the fiber bundle over
space-time with the fiber space  dual of the space of
single-particle quantum states in the system.

 \section{Central On-Mass-Shell Theorem}
The next step is to observe that free massless field equations
in $(A)dS_d$ space can be concisely formulated in terms
of the star product algebra. To this end
one describes the background gravitational field as flat connection
 $\go_0 \in o(d-1,2)$, $R^{A\,B}(\go_0)=0$ with
$
\go_0^{A\,B}\ne0\,,
$
and
$
\omega_0^{A_1\ldots A_{s-1},B_1\ldots B_{s-1}}=0
$
for $s>2$.
In the linearized approximation one sets
$\go(Y|x)=\go_0(Y|x)+\go_1(Y|x)$ where
$\go_1(Y|x)$ describes dynamical fluctuations.
Then the generalization of the free lower spin equations
(\ref{ein})-(\ref{s0}) to the
free equations for massless fields of all spins (plus
constraints on auxiliary fields) is \cite{framed,d}
\be
\label{CMT2}
  R_1 (Y |x)= \half\,\, e_0^a \wedge e_0^b \,\,\f{\p^2}{\p Y^a_i \p
Y^b_j}\,\, \gvep_{ij}\,\, C( Y|x )\Big |_{V_AY^A_i=0}\,,
\ee
\be
\label{CMT}
\widetilde{D}_0 (C) =0\,, \qquad
t_{ij} * C = C*t_{ij}\,,\qquad D(t_{ij})      =0 \,,
\ee
where $R_1$ is the linearized HS field strength (\ref{HSCUR})
and $\widetilde{D}_0 (C)$ is the covariant derivative in the
twisted adjoint representation,
$$ R_1 = d\go + \go_0* \go_1 +\go_1 *\go_0\,, \qquad
\widetilde{D}_0 (C)= d C + \go_0 *C - C*\widetilde{\go}_0\,, $$
\be
\label{til}
\widetilde{f} (Y)=f(\widetilde{Y})\,,\qquad
\widetilde{Y}^A_i=Y^A_i - \f{2}{V^2}\,\, V^A V_B \,Y^B_i\,.
\ee

\section{Nonlinear Construction}

\subsection{General idea}

The form of the equations (\ref{CMT2}), (\ref{CMT})
suggests the idea \cite{Ann} to search
nonlinear HS equations in the ``unfolded" form of
generalized flatness conditions
\be
\label{unf}
 d\omega^\Phi=F^\Phi(\omega)\qquad\qquad
d=dx^n\f{\p}{\p x^n }\,,
\ee
where {$\go$}${}^\Phi(x)$ is a
set of differential forms (including 0-forms)
and the function $F^\Phi(\go)$ contains only wedge
products of {$\go$}${}^\Phi(x)$ and is such that the
consistency condition
\be
\label{cc}
F^\Phi \wedge \f{\gd F^\Omega}{\gd\go^\Phi}=0\,
\ee
is true for any {$\go$}${}^\Phi(x)$. (Once this is the case,
the function $F^\Phi(\omega)$ defines a free
differential algebra \cite{FDA}).

The unfolded form (\ref{unf}) of the field equations
has several nice properties:
\begin{itemize}
\item
It is manifestly invariant under gauge transformations
%\vskip -35pt
\be
\gd\Omega{}^\Phi=d\gvep^\Phi-\gvep^\Omega \f{\gd
F^\Phi}{\gd\go^\Omega}\,,\qquad
\deg
\gvep^\Phi (x)=\deg \go^\Phi (x)-1\,.
\ee
\item
Invariant under diffeomorphisms.
\item
 Interactions: a nonlinear deformation
of~$F^\Phi(\omega)$.
\item
 Degrees of freedom are in the
 0-form fields which form an infinite-dimensional module dual to
the space of single-particle states.
\item
Universality: any dynamical system can be reformulated in the
unfolded form.
\end{itemize}

Originally it was shown by direct inspection that a nonlinear
deformation of the $d=4$ unfolded free massless field equations
(\ref{CMT}) exists in the lowest orders \cite{Ann}. To go beyond
lowest orders some more sophisticated approach was needed. The
useful idea was \cite{more} to find an appropriate generalization
$g^\prime$ of the HS algebra $g$ such that a substitution \be
\label{rest} \go \to W=\go + \go C +\go C^2+\ldots\, \ee into the
$g^\prime$ zero curvature equation $ dW+W\wedge W=0 $ reconstructs
nonlinear HS equations. The key issue is of course to find
restrictions on $W$ that reconstruct (\ref{rest}) in all orders.
The guiding principle  is \cite{gol,d} to preserve $sp(2)$ at the
nonlinear level. Before going into details of the construction let
us mention that the resulting interactions are unique up to field
redefinitions. The only dimensionless coupling constant is the
 YM constant $g^2 = |\Lambda|^{\f{d-2}{2}}
 \kappa^2$ which, however, is artificial
in the classical pure gauge HS theory
because it can be rescaled away just as in the classical
pure Yang-Mills theory.

\subsection{Nonlinear HS Equations}
In \cite{d} it was shown that the appropriate extension
$g\to g^\prime$ is achieved by the doubling of oscillators
$Y^A_i\to (Z^A_i , Y^A_i)$ so that the HS fields extend to
$
\go(Y|x)\to
W(Z,Y|x)\,,$ $ C(Y|x)\to B(Z,Y|x).
$
In addition we introduce the  $S$
connection along $Z^A_i$
that together with $W$ form a noncommutative
connection
\be
 \W = d +W +S\,,\qquad S(Z,Y|x)=dZ^A_i\,S_A^i\,.
\ee
The star product in $g^\prime$ is
\be
\label{star}
(f*g)(Z,Y) = \int dS dT f(Z+S,Y+S)
g(Z-T, Y+T) \exp 2 S_{A}^i T^A_i \,.
\ee
One can see that this is the oscillator algebra
with the nonzero basis relations
$
 [Y^A_i , Y^B_j ]_*=- [Z_i^A , Z_j^B ]_* =
\gvep_{ij}\eta^{AB}\,.
$
It is not however the Moyal star product, being the
normal ordered star product with respect to
 $Z-Y\, \mbox{:}\, Z+Y $ normal ordering because
the left star multiplication by $Z-Y$ and right star
multiplication by $Z+Y$ are equivalent to the usual pointwise
multiplication.

An important property of this star product is that it admits
the Klein operator ${\cal K}$ that generates the automorphism (\ref{til})
\be
 {\cal K} = \exp
{\f{2}{V^2} V_AZ^{Ai} V_B Y^{B}{}_i } \,,\qquad
{\cal K} *f = \widetilde{f}* {\cal K}\,,\qquad {\cal K} *{\cal
K}=1 \,.
\ee

The nonlinear HS field equations can be concisely
formulated in the form \cite{d}
\be
\W *\W =
\half( dZ_A^i dZ^A_i + 4\Lambda^{-1} dz^i dz_{i} B*\K )\qquad \W *
B = B *\widetilde{\W}\,,
\ee
where
$
\widetilde{\W}(dZ
, Z,Y) = \W(\widetilde{dZ}, \widetilde{Z},\widetilde{Y})
$ and
$
dz_i = \f{1}{\sqrt{V^2}} V_B dZ^B_i\,.
$
This system is manifestly gauge invariant under the
gauge transformations
\be
\label{HSsym}
 \delta \W = [\gvep , \W
]_* \,,\qquad \delta B = \gvep * B - B*\widetilde{\gvep}\,
\ee
with an arbitrary gauge parameter
$
\gvep=\gvep (Z,Y|x) \,.
$
One of the most important properties
of the system (\ref{HSsym}) is \cite{d} that it
 admits $sp(2)$ such that its generators $\tau_{ij}$
form a nonlinear deformation of the
$sp(2)$ algebra (\ref{sp2}) and single out the physical
sector of HS fields by the $sp(2)$ invariance
condition $D\tau_{ij} =0$ (followed by factorization of the terms
proportional to $\tau_{ij}$).
The nonlinearly realized $sp(2)$ can be interpreted as a symmetry of a
two-dimensional fuzzy
hyperboloid in the noncommutative space of
$Y^A_i$ and $Z^A_j$. A radius of the fuzzy hyperboloid
depends on  $B(Z,Y|x)$ which is the generating
function for the Weyl 0-forms.

To analyse the HS field equations perturbatively one sets
\be
\label{vacu}
 W=W_0 +W_1\,,\qquad S=  dZ^A_i Z_A^i+S_1 \,,\qquad B=B_1\,,
\ee
where
$
W_0 = \half \go_0^{AB} (x) Y^i_A Y_{iB}
$
with $\go_0^{AB}(x)$ describing
background $AdS_d$ gravitational field. It is not hard to see
that central on-mass-shell theorem (\ref{CMT}) is reproduced
in the lowest order \cite{d}.

Let us note that the form of the noncommutative connection $S$ in
(\ref{vacu}) implies that, because of the first term in $S$,
 the HS symmetry (\ref{HSsym}) is
spontaneously broken down to the HS symmetry with $Z$-independent
parameters $\gvep (Y|x)$ (The $Z$-dependent components in
$\gvep (Z,Y|x)$ are used to gauge fix the
noncommutative $Z-$connection $S$). Because of the $B$-dependent term
in (\ref{HSsym}), the leftover HS symmetries with the HS gauge parameters
$\gvep (Y|x)$ acquire $B$-dependent nonlinear corrections. As a
result, HS gauge symmetries in the nonlinear HS theory are different
from the Yang-Mills gauging of the global
HS symmetry of a free theory one starts with.

\section{Singletons in any Dimension}

The simplest HS algebra
 $hu(1|2\!\!:\!\![d\!-\!1,2])$ admits the
fermionic generalization $hu(1|(1,2)\!\!:\!\![d\!-\!1,2])$ \be
sp(2)\to osp(1,2)\qquad Y^A_i \to (Y^A_i, \phi^A )
  \,,\qquad \{\phi^A , \phi^B \} =
-2\eta^{AB}\,.
\ee
The  Fock-type modules of $hu(1|2\!\!:\!\![M,2])$
and $hu(1|(1,2)\!\!:~\!\![M,2])$ describe
 massless scalar $S_{M}$ and spinor $F_{M}$ in $M$
dimensions \cite{sing}
(closely related analysis of $M$-dimensional
field equations in terms of conformal algebra
and dual $sp(2)$ and $osp(1,2)$ was given in
\cite{mar,bars}).
This gives  realization of
the HS algebras as conformal HS algebras acting on the
scalar and spinor conformal fields (i.e., singletons)
in $M$ dimensions.

The following generalization of the $4d$ Flato-Fronsdal theorem
\cite{FF} takes place \cite{sing}:
\be
S_{d-1}\otimes S_{d-1} =\sum_{s=0 }^\infty \oplus
%diagrammaodnoryadnaya nad ney s sprava ot nee
\,\, { \begin{picture}(80,10)
%palka s
{\linethickness{.250mm}
\put(00,00){\line(1,0){80}}%
\put(00,10){\line(1,0){80}}%
\put(00,00){\line(0,1){10}}%
\put(10,00.0){\line(0,1){10}} \put(20,00.0){\line(0,1){10}}
\put(30,00.0){\line(0,1){10}} \put(40,00.0){\line(0,1){10}}
\put(50,00.0){\line(0,1){10}} \put(60,00.0){\line(0,1){10}}
\put(70,00.0){\line(0,1){10}} \put(80,00.0){\line(0,1){10}} }
\put(22,12.){  ${s}$} \end{picture}}\quad m=0\quad \mbox{{
bosons in} } AdS_d\,
\ee
(note that related statements were discussed in
\cite{SS})
\be
 F_{d-1}\otimes S_{d-1}
=\sum_{s=\half }^\infty \oplus\,\,
%diagramma odnoryadnaya nad ney
 { \begin{picture}(80,10)
%palka s
{\linethickness{.250mm}
\put(00,00){\line(1,0){80}}%
\put(00,10){\line(1,0){80}}%
\put(00,00){\line(0,1){10}}%
\put(10,00.0){\line(0,1){10}} \put(20,00.0){\line(0,1){10}}
\put(30,00.0){\line(0,1){10}} \put(40,00.0){\line(0,1){10}}
\put(50,00.0){\line(0,1){10}} \put(60,00.0){\line(0,1){10}}
\put(70,00.0){\line(0,1){10}} \put(80,00.0){\line(0,1){10}} }
\put(2,12){ \small ${s-1/2}$}\end{picture}}\quad  m=0\quad
\mbox{{ fermions in }} AdS_d
\ee
\\
$$
  F_{d-1}\otimes F_{d-1}
=\sum_{p,q }\oplus\,\,\,\,\,\,\,
 { \begin{picture}(80,20)(0,20)
%palka s
{\linethickness{.250mm}
\put(00,50){\line(1,0){80}}%
\put(00,00){\line(1,0){10}}%
\put(00,10){\line(1,0){10}}%
\put(00,20){\line(1,0){10}}%
\put(00,30){\line(1,0){10}}%
\put(00,40){\line(1,0){80}}%
\put(00,0){\line(0,1){50}}%
\put(10,00.0){\line(0,1){50}} \put(20,40.0){\line(0,1){10}}
\put(30,40.0){\line(0,1){10}} \put(40,40.0){\line(0,1){10}}
\put(50,40.0){\line(0,1){10}} \put(60,40.0){\line(0,1){10}}
\put(70,40.0){\line(0,1){10}} \put(80,40.0){\line(0,1){10}} }
\put(22,62){ \small ${p+1}$} \put(-20,22){ \small ${q}$}
\end{picture}} \quad m=0\quad \mbox{{ bosons in} } AdS_d\,
\nonumber
$$
\be
\oplus m>0\quad \mbox{ { antisymmetric tensors}}
\ee

These results show precise
matching between spectra of gauge fields in HS theories
and appropriate UIRs
of HS algebras, indicating the existence of HS gauge theories
with fermions and mixed symmetry fields \cite{sing}. Moreover,
HS superalgebras exist in any $d$ \cite{sing}. There is no
contradiction here with the absence of usual supersymmetries in
higher dimensions
because HS superalgebras contain usual finite-dimensional
subsuperalgebras  only for some lower dimensions like
$d$ $=3,4,5$.

\section{Conclusions}
The main conclusion is that nonlinear HS theories
exist in any dimension. Note that HS gauge symmetries in the nonlinear HS
theory differ from the Yang-Mills gauging of the global
HS symmetry of a free theory one starts with by HS field strength
dependent nonlinear corrections
resulting from the partial gauge fixing of spontaneously broken HS
symmetries in the extended noncommutative space.

The HS geometry is that of the
fuzzy hyperboloid in the auxiliary (fiber) noncommutative space.
Its radius depends on the Weyl 0-forms which
take values in
the infinite-dimensional module dual to the space of single-particle
states in the system.

\section*{Acknowledgments}
This research was supported in part by grants
INTAS No.03-51-6346, RFBR No.02-02-17067 and LSS No.1578.2003-2.

\end{document}